# Quality of Heusler Single Crystals Examined by Depth Dependent Positron Annihilation Techniques

C Hugenschmidt[1,2*], A Bauer[1], P Böni[1], H Ceeh[1], S W H Eijt[3], T Gigl[1,2],
C Pfleiderer[1], C Piochacz[1,2], A Neubauer[1], M Reiner[1,2], H Schut[3], and J Weber[1]

[1]Physik Department E21, Technische Universität München, Garching, Germany
[2]Heinz Maier-Leibnitz Zentrum (MLZ), Technische Universität München, Garching, Germany
[3]Department of Radiation Science and Technology RST Faculty of Applied Sciences, Delft University of Technology, the Netherlands

E-mail: christoph.hugenschmidt@frm2.tum.de

**Abstract**. Heusler compounds exhibit a wide range of different electronic ground states and are hence expected to be applicable as functional materials in novel electronic and spintronic devices. Since the growth of large and defect-free Heusler crystals is still challenging, single crystals of $Fe_2TiSn$ and $Cu_2MnAl$ were grown by the optical floating zone technique. Two positron annihilation techniques – Angular Correlation of Annihilation Radiation (ACAR) and Doppler Broadening Spectroscopy (DBS) – were applied in order to study both, the electronic structure and lattice defects. Recently, we succeeded to observe clearly the anisotropy of the Fermi surface of $Cu_2MnAl$, whereas the spectra of $Fe_2TiSn$ were disturbed by foreign phases. In order to estimate the defect concentration in different samples of Heusler compounds the positron diffusion length was determined by DBS using a monoenergetic positron beam.

## 1. Introduction

Heusler crystals comprise a large class of ternary intermetallic compounds with the $L2_1$ structure type and with the formula $X_2YZ$ where X, Y are transition metals, and Z is a non-magnetic or a non-metallic element. The family of Heusler compounds exhibits an extraordinary wide range of different electronic ground states with metallic, semiconducting, insulating, half-metallic, ferromagnetic or superconducting behavior [1]. For novel electronic and spintronic applications great efforts are undertaken to benefit from the electronic properties of this class of materials. For this reason, it is a major task to develop new functional materials based on Heusler compounds. In order to enable the investigation of basic properties such as the electronic structure the availability of Heusler single-crystals is of major importance. However, the growth of large and defect-free single-crystals of this class of ternary alloys remains still challenging.

Within this contribution, we focus on the ternary systems $Cu_2MnAl$ and $Fe_2TiSn$. $Cu_2MnAl$ is the Heusler system per se: In 1903, Fritz Heusler discovered that this compound behaves like a magnet, although none of its constituent elements is ferromagnetic by itself [2]. The Curie temperature of $Cu_2MnAl$ is found to be $T_C = 622$ K. For $Fe_2TiSn$ various properties are theoretically predicted such as a non-magnetic ground state, a pseudogap and a very low density of states at the Fermi level [3,4].

---
[*] To whom any correspondence should be addressed.

For the present project, we applied the 2D-ACAR technique in order to study the anisotropy of the Fermi surface. Although positron trapping in defects disturbs the measured electron momentum distribution Dugdale et al. could show that the radial anisotropy of 2D-ACAR spectra can still be recovered since the defect contribution is mainly isotropic [5]. However, both the size of the Heusler single crystals and a low amount of (open-volume) defects are important for the investigation of the electronic structure. For this reason, $Fe_2TiSn$ crystals and particularly large single-crystals of $Cu_2MnAl$ were produced using the optical floating zone technique [6,7]. The quality of the crystalline samples, i.e. the amount of defects, was examined with DBS using a monoenergetic positron beam. Hence, the positron diffusion length, which is related with the concentration of positron trapping sites, could be determined by the depth dependent measurement of the annihilation line shape parameter S. The application of a simple trapping model allows us to estimate the order of magnitude of the vacancy concentration more quantitatively in different samples of the Heusler compounds.

## 2. Experimental

*2.1. Sample preparation*

Large single-crystals of the Heusler compounds were grown by means of optical floating zone technique under ultra-high vacuum conditions. The heat is generated by high-power lamps in the focal point of elliptical mirrors yielding a temperature of more than 2000°C. Due to this crucible-free technique contamination with foreign atoms can be avoided. Starting from high-purity polycrystals single crystalline $Fe_2TiSn$ samples were grown as well as single crystals of $Cu_2MnAl$. Using the optical floating zone technique large single crystals with a diameter of more than 10 mm could be obtained. Details of this technique and in particular on the single crystal growth of $Cu_2MnAl$ can be found in reference [7]. Before each measurement the $Fe_2TiSn$ samples were soaked in one molar hydrochloric acid for 20 minutes, cleaned with isopropanol and quickly transferred into the vacuum chamber. After the cleaning procedure a residual pressure of less than $10^{-2}$ mbar was reached within less than 60 seconds.

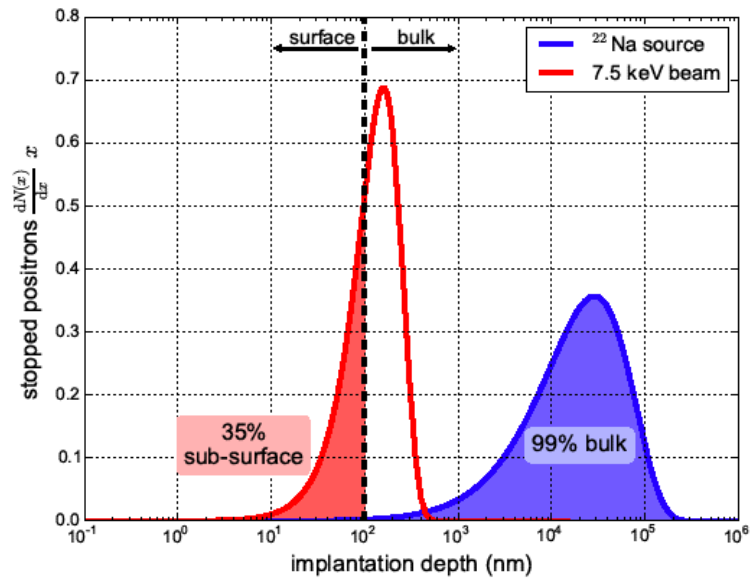

**Figure 1.** The implantation profile of positrons: The implantation of 7.5 keV positrons leads to mean implantation depth in $Fe_2TiSn$ of nearly 200 nm. Using a $^{22}$Na source information of the bulk is obtained since only 1% of the positrons probe the sub-surface region.

*2.2. Measurements*

Two dimensional projections of the Two-Photon Momentum Distribution (TPMD) were measured using the 2D-ACAR technique. First, the TPMD of an oriented $Fe_2TiSn$ single crystal with the integration axis along [100] was measured at the positron beam facility POSH at the Reactor Instituut of the Delft University of Technology [8,9]. The sample was aligned with the aid of single crystal X-ray diffraction. In order to compare the spectra in the sub-surface region and in the bulk ACAR measurements were performed with both the positron beam at 7.5 keV and a $^{22}Na$ source at POSH. The depth dependent positron densities shown in Figure 1 were calculated using Makhovian implantation profiles. For the 2D-ACAR spectrum obtained with the beam $7.3 \cdot 10^7$ counts were collected within 5.5 days. Using the relatively weak $^{22}Na$ positron source $2 \cdot 10^7$ counts were collected within 12 days. In another experiment, ACAR spectra of $Cu_2MnAl$ were recorded with the new 2D-ACAR spectrometer at the Technische Universität München (TUM). This set-up uses an optimized static magnetic field to guide the positrons from a $^{22}Na$ source onto the sample [10]. During six days $1.5 \cdot 10^8$ counts were recorded for each 2D-ACAR spectrum.

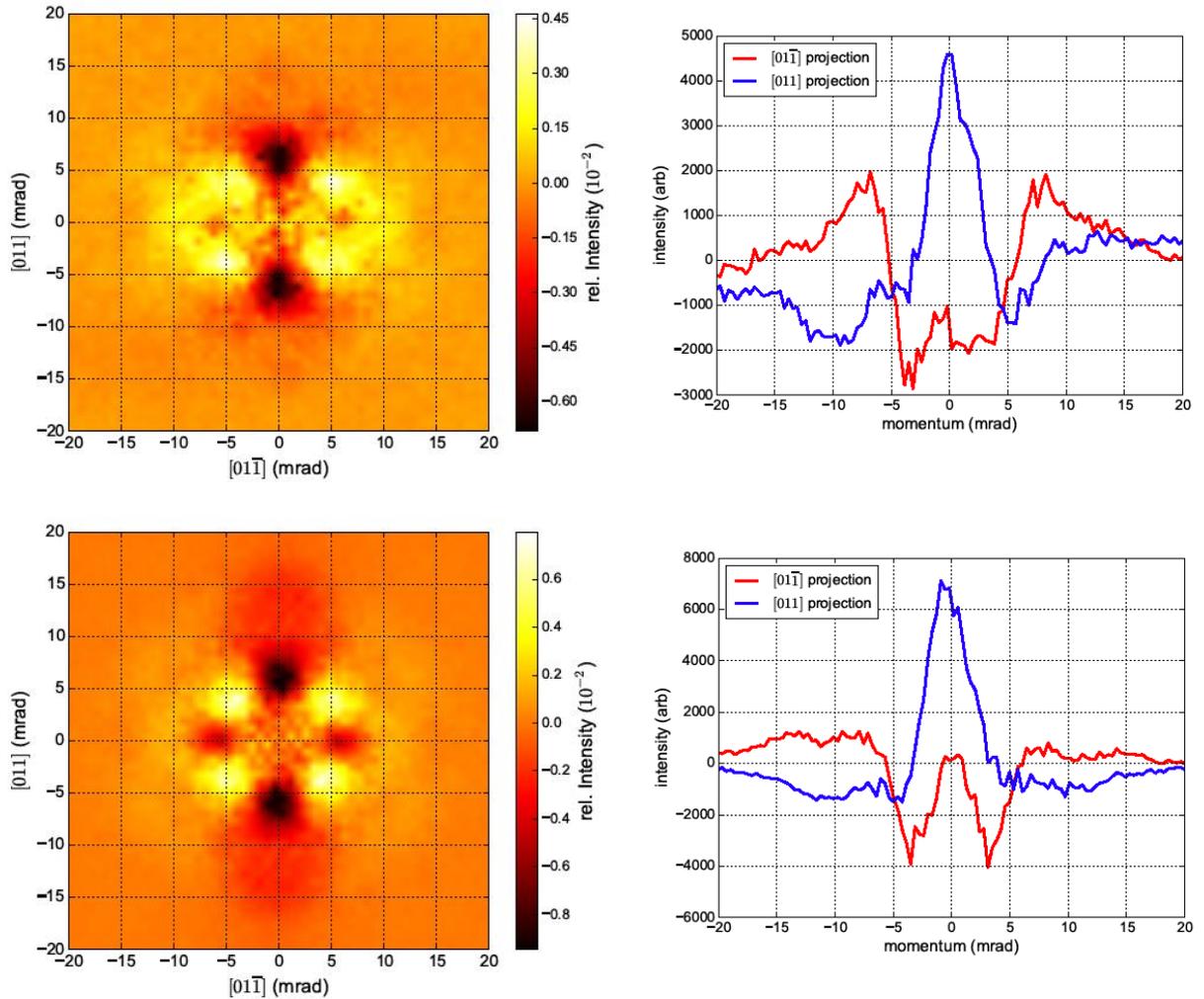

**Figure 2.** The anisotropy of the 2D-ACAR spectra shows the projection along the [100] direction of $Fe_2TiSn$ (left) and 1D-projections along high-symmetry directions (right). The spectra are recorded with a 7.5 keV positron beam (upper figures) and with a $^{22}Na$ source (below). The deviation from the expected 4-fold symmetry is clearly observed in both measurements.

The Doppler broadening measurements were performed at the coincident Doppler broadening spectrometer [11,12] at the positron source NEPOMUC [13,14] at the high flux neutron source FRM II of the TUM. Two different samples of $Fe_2TiSn$ and one $Cu_2MnAl$ single crystal were measured with depth dependent DBS. The energy of the positron beam was varied in the range of 1–30 keV, and the measurement time was set to one minute for each energy value.

**3. Results and Discussion**

*3.1. ACAR measurements*

First we measured 2D-ACAR spectra with the projection direction along [100] for the $Fe_2TiSn$ sample close to the surface (7.5 keV beam) and in the bulk ($^{22}$Na positron source). The recorded anisotropy of the 2D-ACAR spectra is shown in Figure 2. In the measured spectrum the anisotropy of the TPMD of an oriented single crystal should lead to a pattern according to the symmetry of the integration direction. However, a clear deviation from the expected 4-fold symmetry is observed for both measurements. This becomes even more apparent in the 1D-projections along [011] and [01-1] which should be the same for a pure 4-fold symmetry.

A semi-quantitative analysis by means of the auto-correlation around the rotational axis reveals that this effect is more dominant at the surface than in the bulk of the sample (see figure 3). The surface signal, which was obtained with a positron implantation energy of 7.5 keV shows a weaker 4-fold symmetric contribution than the bulk signal, which was obtained by means of conventional 2D-ACAR using the $^{22}$Na positron source.

One possible explanation for this observation could be the presence of a foreign iron rich phase. It is known from SEM and EDX that $Fe_{68}Ti_{25}Sn_7$ can be formed predominantly at grain boundaries. Since positrons can be trapped at grain boundaries this foreign phase could contribute to the pattern observed in the 2D-ACAR spectrum disturbing the 4-fold symmetry. In addition, structural vacancies are assumed to be present in the sample. Positron annihilation in such trapping sites would lead to a reduced anisotropy in the TPMD.

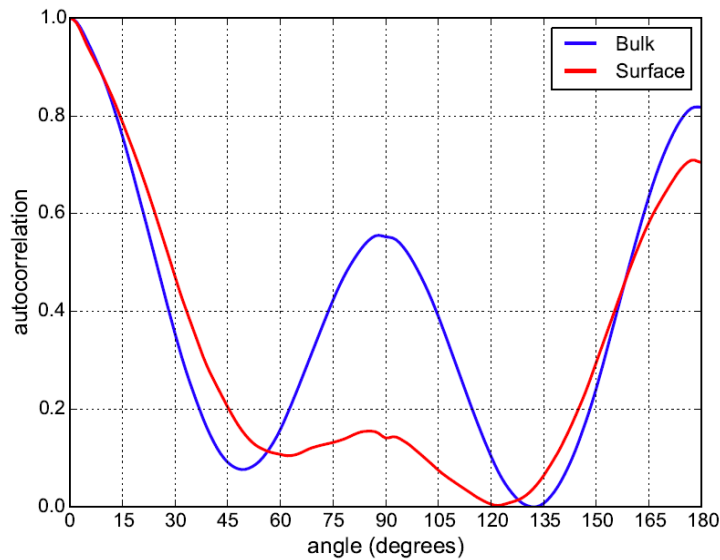

**Figure 3.** Auto-correlation of the anisotropy as a function of the rotation angle in $Fe_2TiSn$. The rotational axis is by minimizing the moment of inertia in the raw 2D-ACAR data. For 4-fold (2-fold) symmetric data maxima are expected every 90 (180) degrees. The bulk data obtained with the $^{22}$Na source exhibits partially a 4-fold symmetry whereas the signal, which originates from the near surface area of the sample (positron beam measurement), shows a dominant 2-fold symmetric contribution.

The anisotropy recorded for $Cu_2MnAl$ can clearly be seen with high signal to noise in Figure 4 for both projections along [100] and [110]. As expected the according 4-fold and 2-fold symmetries are obvious for both projections. The symmetries of the recorded spectra for $Cu_2MnAl$ can also be displayed by analyzing the data with the auto-correlation of the anisotropy as a function of the rotation angle (see Figure 5). It should be mentioned that the color scale in Figures 2 and 4 is the same but the range of the anisotropy of $Cu_2MnAl$ is about two orders of magnitude higher than for $Fe_2TiSn$. Since the ACAR spectra obtained for $Cu_2MnAl$ show a much more pronounced anisotropy than for $Fe_2TiSn$ it is expected that this single crystal contains much less positron trapping sites at vacancies than the $Fe_2TiSn$ samples. Due to these promising results spin-resolved ACAR measurements using this $Cu_2MnAl$ single crystal become possible for the first time.

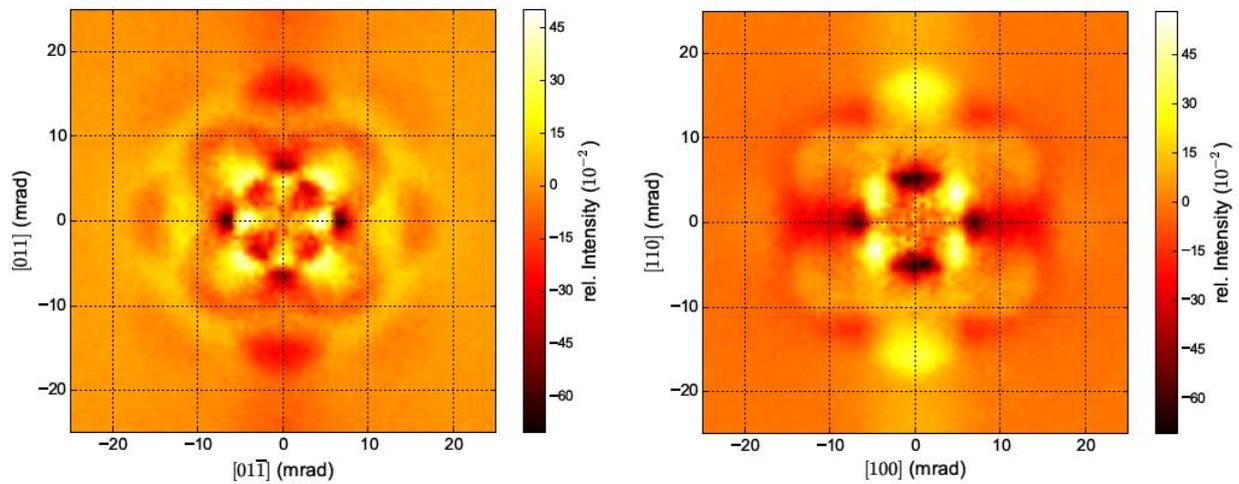

**Figure 4.** Anisotropy of the 2D-ACAR spectrum of $Cu_2MnAl$: The projections along [100] (left) and [110] (right) show clearly the expected 4-fold and 2-fold symmetries, respectively.

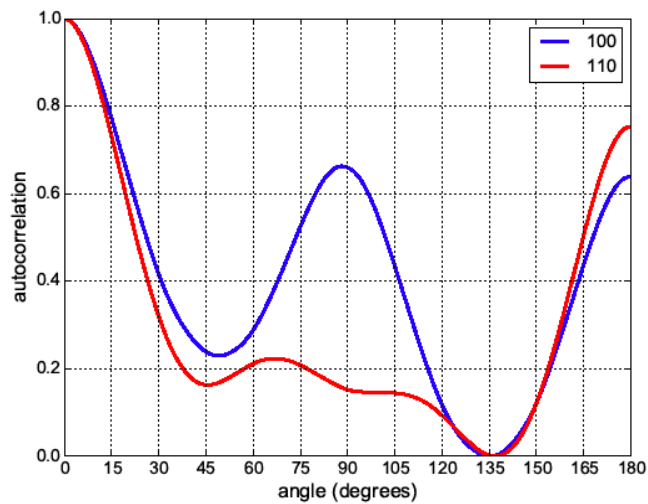

**Figure 5.** Auto-correlation of the anisotropy as a function of the rotation angle in $Cu_2MnAl$. The rotational axis is by minimizing the moment of inertia in the raw 2D-ACAR data. As expected the projections along [100] and [110] show clearly the 4-fold and 2-fold symmetries, respectively.

*3.2. Depth dependent DBS*

Depth dependent DBS measurements were performed in order to estimate the defect concentration in the samples. The slope of the S-parameter (defined in the usual way, see e.g. [15] and references therein) is shown as a function of the positron implantation energy for two different $Fe_2TiSn$ samples and for $Cu_2MnAl$ in Figure 6. The high value at the surface (~1keV) is attributed to the annihilation of positrons in a surface state and from positronium. With increasing positron implantation energy the S parameter decreases until a saturation value is reached in the bulk, i.e. no positrons diffuse back to the surface. Assuming that the measured S parameter is a superposition of an S value at the surface and a bulk value we can determine the positron diffusion length using the program VEPFIT [16]. The fit results are plotted as red lines in Figure 6.

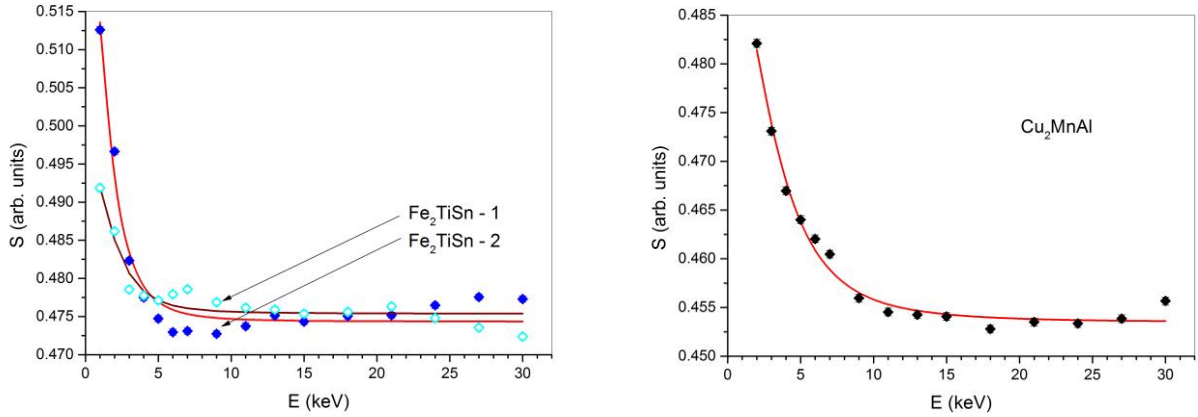

**Figure 6.** Depth dependent Doppler broadening spectroscopy: The S-parameter as a function of positron implantation energy for two different $Fe_2TiSn$ samples (left) and $Cu_2MnAl$ (right) are shown with the VEPFIT results (red lines). In $Cu_2MnAl$ the determined positron diffusion length is roughly a factor of three higher than in $Fe_2TiSn$.

The positron diffusion length $L_+$ was determined to 13 and 19 nm in the $Fe_2TiSn$ samples (the latter belongs to the sample measured with ACAR, see Figure 2), and to 43 nm in the $Cu_2MnAl$ crystal. A simple model assuming only vacancies as positron trapping sites allows the estimation of the vacancy concentration $c_v$ in the samples using the equation (see e.g. [16])

$$c_v = \left(\frac{D_+}{L_+^2} - \frac{1}{\tau_b}\right)/\mu$$

with the positron diffusion coefficient $D_+$, the positron trapping rate $\mu$, and the bulk positron lifetime $\tau_b$. Since values for the Heusler compounds $Fe_2TiSn$ and $Cu_2MnAl$ are not known reasonable values are taken as following: $D_+ = 0.5(1) cm^2/s$, $\mu=4(1)10^{14} s^{-1}$ and $\tau_b=120(20)$ ps. It is noteworthy that the error of $c_v$ is governed by the errors of $D_+$ and $\mu$; the error of $\tau_b$ is not relevant. Hence, the vacancy concentration $c_v$ can be estimated to be in the range of $1.6 \cdot 10^{-4}$ to $1.6 \cdot 10^{-3}$ for the $Fe_2TiSn$ samples whereas it amounts to $c_v = 2.0 \cdot 10^{-5} - 9.7 \cdot 10^{-5}$ in the $Cu_2MnAl$ single crystal. Consequently, the vacancy concentration is approximately one order of magnitude less in the $Cu_2MnAl$ single crystal compared to the $Fe_2TiSn$ samples leading to a much stronger signal in the anisotropy of the TPMD measured with 2D-ACAR.

## 4. Summary and Outlook

The preparation of single crystalline samples with as small as possible defect concentration is crucial for (spin-polarized) ACAR experiments. From the reduced or even missing anisotropy in the ACAR spectra it is difficult to distinguish between structural defects and disorder, which to some extent is inherent to any Heusler system. In this context, depth dependent DBS is a valuable tool for sample characterization to evaluate high vacancy concentrations even up to the percent range. The grown $Cu_2MnAl$ single crystal showed a high quality leading to a pronounced anisotropy in the 2D-ACAR spectra. For this reason, spin-polarized ACAR measurements and theoretical calculations are underway in order to study the anisotropy of the Fermi surface for each spin channel in this ferromagnetic Heusler compound.

**Acknowledgement**
This research project was funded by the Deutsche Forschungsgemeinschaft (DFG) within the Transregional Collaborative Research Center TRR80 "From electronic correlations to functionality" and supported by the European Commission under the 7th Framework Programme through the 'Research Infrastructures' action of the 'Capacities' Programme, Contract No: CP-CSA_INFRA-2008-1.1.1 Number 226507-NMI3. Financial support by the BMBF (project nos. 05KI0WOB and 05K13WO1) is gratefully acknowledged.

**References**
[1] T. Graf, C. Felser, S. S. Parkin, Progress in Solid State Chemistry 39 (2011) 1
[2] F. Heusler, Verhandlungen der Deutschen Physikalischen Gesellschaft 5 (1903) 1
[3] A. Ślebarski, Phys. Rev. B 62 (2000) 3296
[4] A. Ślebarski, J. Phys. D: Appl. Phys. 39 (2006) 856
[5] S. B. Dugdale, , and J. Laverock, J. Phys. Conf. Ser. 505 (2014) 012046
[6] A. Neubauer. Ph.D. thesis, Technische Universität München (2011)
[7] A. Neubauer, F. Jonietz, M. Meven, R. Georgii, G. Brandl, G. Behr, P. Böni, and C. Pfleiderer, Nuclear Instruments and Methods in Physics Research Section A, 688, (2012) 66
[8] C.V. Falub, P.E. Mijnarends, S.W.H. Eijt, M.A. van Huis, A. van Veen, and H. Schut, Phys. Rev. B 66 (2002) 075426.
[9] W. Al-Sawai, B. Barbiellini, Y. Sakurai, M. Itou, P. E. Mijnarends, R. S. Markiewicz, S. Kaprzyk, S. Wakimoto, M. Fujita, S. Basak, H. Lin, Yung Jui Wang, S.W.H. Eijt, H. Schut, K. Yamada, and A. Bansil, Phys. Rev. B 85(2012) 115109
[10] H. Ceeh, J. A. Weber, M. Leitner, P. Böni, and C. Hugenschmidt, Rev. Sci. Instr. 84 (2013) 043905
[11] M. Stadlbauer, C. Hugenschmidt, and K. Schreckenbach, Appl. Surf. Sci. 255 (2008) 136
[12] M. Reiner, P. Pikart, and C. Hugenschmidt, J. Phys. Conf. Ser. 443(2013) 012071
[13] C. Hugenschmidt, B. Löwe, J. Mayer, C. Piochacz, P. Pikart, R. Repper, M. Stadlbauer, and K. Schreckenbach, Nucl. Instr. Meth. A 593 (2008) 616
[14] C. Hugenschmidt, H. Ceeh, T. Gigl, F. Lippert, C. Piochacz, M. Reiner, K. Schreckenbach, S. Vohburger, J. Weber, and S. Zimnik, J. Phys. Conf. Ser. 505 (2014) 012029
[15] C. Hugenschmidt, N. Qi, M. Stadlbauer, and K. Schreckenbach, Phys. Rev. B 80 (2009) 224203
[16] A. van Veen and H. Schut and J. de Vries and R. A. Hakvoort and M. R. IJpma, AIP Conf. Proc. 218 (1991) 171